\documentclass[
 aip,
 adv,
 amsmath,amssymb,
 preprint
]{revtex4-2}

\usepackage[utf8]{inputenc}
\usepackage[T1]{fontenc}
\usepackage[english]{babel}
\usepackage{graphicx}
\usepackage{dcolumn}
\usepackage{bm}

\begin{document}

\title{Importance of nuclear quantum effects on the structure of supercooled water around its liquid--liquid critical point}

\author{M. Beerbaum}
\affiliation{Center for Advanced Systems Understanding (CASUS), Conrad-Schiedt-Stra{\ss}e 20, 02826 G\"orlitz}
\affiliation{Helmholtz Zentrum Dresden-Rossendorf, Bautzner Landstra{\ss}e 400, 01328 Dresden, Germany}

\author{J. Heske}
\affiliation{Chair of Theoretical Chemistry and Center for Sustainable Systems Design, Paderborn University, Warburger Str. 100, D-33098 Paderborn, Germany}

\author{J. Gujt}
\affiliation{Chair of Theoretical Chemistry and Center for Sustainable Systems Design, Paderborn University, Warburger Str. 100, D-33098 Paderborn, Germany}

\author{Thomas D. K\"uhne}
\email{t.kuehne@hzdr.de}
\affiliation{Center for Advanced Systems Understanding (CASUS), Conrad-Schiedt-Stra{\ss}e 20, 02826 G\"orlitz}
\affiliation{Helmholtz Zentrum Dresden-Rossendorf, Bautzner Landstra{\ss}e 400, 01328 Dresden, Germany}
\affiliation{Institute of Artificial Intelligence, Technische Universit\"at Dresden, Helmholtzstra{\ss}e 10, 01069 Dresden, Germany}

\date{\today}

\begin{abstract}
Supercooled water is expected to exhibit a liquid--liquid phase transition between low- and high-density liquid states, possibly terminating in a liquid--liquid critical point in the experimentally difficult no man's land. Because the hydrogen atoms are light, nuclear quantum effects (NQE) may alter the structural signatures used to identify this transition. Here, we compare classical molecular dynamics and path-integral molecular dynamics simulations of a flexible q-TIP4P/F-like water model in the deeply supercooled regime. The classical simulations show a pronounced density change at 180 K between 180 and 220 MPa, whereas the path-integral simulations exhibit a smoother pressure dependence. Radial distribution functions and bond-order parameters show that NQE broaden pair correlations, reduce the tetrahedral order of the first hydration shell, and slightly increase the Steinhardt $Q_6$ parameter. These results demonstrate that NQE modify both low- and high-density liquid structures and therefore need to be included when interpreting structural signatures of the liquid--liquid transition in supercooled water.
\end{abstract}

\maketitle

\section{Introduction}
\label{sec:introduction}

Water is one of the most extensively studied molecular liquids, yet the microscopic origin of many of its anomalies remains incompletely understood \cite{Ball2008}. In the supercooled regime, thermodynamic response functions such as the isobaric heat capacity and the isothermal compressibility increase rapidly below the homogeneous ice-nucleation temperature $T_\mathrm{H}$ \cite{Debenedetti2003,Stanley2000}. A widely discussed interpretation attributes this behavior to a metastable first-order liquid--liquid phase transition (LLPT) between low-density liquid (LDL) and high-density liquid (HDL) water, ending in a liquid--liquid critical point (LLCP) \cite{Stanley1998}.

Direct experimental access to this region is difficult because rapid crystallization occurs below $T_\mathrm{H}$, which is why it is often referred to as water's no man's land. Experimental strategies to circumvent crystallization include aqueous solutions \cite{Mishima2005,Mishima2007,Murata2012,Murata2013,Mallamace2012}, confined water \cite{Wang2015,Mallamace2012,Liu2005,Mallamace2007}, and nanodroplet experiments \cite{Sellberg2014,Manka2012}. In parallel, molecular simulations have provided detailed microscopic information on the metastable liquid. Poole \textit{et al.} first proposed a LLPT for the ST2 water model \cite{Poole1992}, and Palmer \textit{et al.} later estimated the corresponding LLCP at 230 K and 240 MPa \cite{Palmer2015}. For TIP4P/2005 water, Abascal and Vega reported a LLCP at 193 K and 135 MPa \cite{Abascal2010}, while subsequent studies located it near 182 K and 158--162 MPa \cite{Sumi2013,Yagasaki2014}. The existence and interpretation of the LLPT remain debated, with alternative explanations emphasizing finite-size effects and local ice-like ordering \cite{Limmer2013,Limmer2015,Overduin2013,Overduin2015,Ferdinand1969,Fisher1972}.

Most earlier simulations of the LLPT used classical nuclei. This approximation is not obviously benign for water, where proton delocalization and zero-point motion can influence hydrogen-bond structure and dynamics. Previous PIMD/RPMD and related simulations of ambient liquid water have shown that NQE soften the liquid structure and modify translational, vibrational, and hydrogen-bond dynamics \cite{Spura2014,Ojha2018JCP,Clark2019}. These findings motivate testing whether the same quantum-nuclear softening changes the sharper structural signatures expected near the LLCP. In this work, we assess the importance of NQE near the expected LLCP by comparing classical molecular dynamics (MD) with path-integral molecular dynamics (PIMD) simulations using the flexible q-TIP4P/F-like model of Spura \textit{et al.} \cite{Spura2014}. We analyze densities, radial distribution functions (RDFs), the local tetrahedral order parameter $q_4$ \cite{Chau_Hardwick,Errington2001}, and the Steinhardt order parameter $Q_6$ \cite{Steinhardt1983,Lechner2008} to determine how NQE affect the structural distinction between LDL-like and HDL-like states.

\section{Methods}
\label{sec:methods}

\subsection{Path-integral molecular dynamics}
\label{sec:pimd}

Equilibrium PIMD exploits the classical isomorphism of discretized quantum statistical mechanics \cite{Chandler1981} and maps each quantum nucleus onto a classical ring polymer of $P$ beads connected by harmonic springs \cite{Parrinello1984}. In the limit $P \to \infty$, this representation samples the exact quantum-mechanical equilibrium distribution. Up to normalization constants, the discretized canonical partition function can be written as
\begin{equation}
  Z_P \propto \int d\mathbf{p}\, d\mathbf{r}\,
  \exp[-\beta_P H_P(\mathbf{p},\mathbf{r})],
\end{equation}
where $\beta_P=\beta/P$. The ring-polymer Hamiltonian is
\begin{equation}
  H_P =
  \sum_{k=1}^{P}
  \left[
    \sum_{i=1}^{N}
    \left(
      \frac{\mathbf{p}_{i,k}^{2}}{2m_i}
      + \frac{1}{2}m_i\omega_P^2
      |\mathbf{r}_{i,k}-\mathbf{r}_{i,k+1}|^2
    \right)
    + V(\mathbf{r}_{1,k},\ldots,\mathbf{r}_{N,k})
  \right],
\end{equation}
with $\omega_P=P/(\beta\hbar)$ and cyclic boundary conditions in the bead index. Because the intermolecular interactions have to be evaluated for the ring-polymer representation, PIMD is more expensive than classical MD. We reduce this cost using the ring-polymer contraction approach \cite{Markland2008}, in which the expensive long-range electrostatic contribution is evaluated on the centroid while the intramolecular terms retain the full bead resolution. Related auxiliary-potential contraction schemes have enabled ab initio PIMD of liquid water at nearly classical simulation cost \cite{John2016}.

\subsection{Structural order parameters}
\label{sec:order_parameters}

Local tetrahedrality is quantified using the bond-orientational order parameter introduced by Chau and Hardwick \cite{Chau_Hardwick} and later used by Errington and Debenedetti \cite{Errington2001},
\begin{equation}
  q_i = 1 - \frac{3}{8}
  \sum_{1\le j<k\le 4}
  \left(\cos\,\theta_{jik}+\frac{1}{3}\right)^2 ,
\end{equation}
where $\theta_{jik}$ is the angle formed by the vectors from molecule $i$ to two of its four nearest neighbors. A perfectly tetrahedral environment gives $q_i=1$, while uncorrelated configurations average to zero. The system-averaged tetrahedral order is
\begin{equation}
  q_4 = N^{-1}\sum\nolimits_{i=1}^{N} q_i .
\end{equation}

To characterize structural order beyond the first hydration shell, we also compute the Steinhardt order parameter \cite{Steinhardt1983,Lechner2008}
\begin{equation}
  Q_{l,i} =
  \sqrt{
    \frac{4\pi}{2l+1}
    \sum_{m=-l}^{l}
    \left|\bar{Y}_{lm}(i)\right|^2
  },
\end{equation}
where $\bar{Y}_{lm}(i)$ is the average of the spherical harmonic $Y_{lm}$ over the bond vectors connecting molecule $i$ with its twelve nearest water molecules. We focus on $l=6$ and use
\begin{equation}
  Q_6 = \frac{1}{N}\sum_{i=1}^{N}Q_{6,i}.
\end{equation}
For an uncorrelated set of $k$ bond directions, $Q_6^{\mathrm{ig}}=1/\sqrt{k}$, which gives $Q_6^{\mathrm{ig}}\approx 0.289$ for $k=12$ \cite{Q6_ideal_gas}. Values for crystalline reference structures are larger, for example 0.574 for fcc, 0.484 for hcp, and 0.510 for bcc environments \cite{Q6_Mixing_effects_in_glass_forming_lennard-jones_mixtures}.

\subsection{Simulation details}
\label{sec:simulation_details}

All simulations were performed in the isothermal--isobaric ensemble using periodic supercells containing 125 water molecules. Interactions were described with the q-TIP4P/F-like flexible water model of Spura \textit{et al.} \cite{Spura2014}, which was parametrized by force matching to DFT-based TPSS-D3 reference calculations \cite{Koster2016}. This model is based on the q-TIP4P/F model of Habershon \textit{et al.} \cite{Habershon2009}, which in turn was derived from the rigid TIP4P/2005 model of Abascal and Vega \cite{Abascal2005}. Unlike rigid TIP4P/2005, the q-TIP4P/F-like model includes intramolecular harmonic bending and anharmonic Morse stretching terms and is therefore suitable for simulations with quantum nuclei.

Each trajectory was pre-equilibrated for 0.5 ns in the canonical ensemble, followed by 3 ns of equilibration in the isothermal--isobaric ensemble. Production trajectories of 10 ns were then used for analysis. Temperature was controlled with an Andersen thermostat \cite{Andersen1980}, and pressure with a Berendsen barostat \cite{Berendsen1984}. Multiple-time-step integration \cite{Tuckerman1992} used time steps of 0.25 fs for intermolecular forces and 0.5 fs for intramolecular forces. The temperature was varied from 140 to 240 K in steps of 20 K, and the pressure from 100 to 260 MPa in steps of 40 MPa. PIMD simulations used $P=64$ beads; otherwise identical classical reference simulations were carried out with $P=1$.

\section{Results and Discussion}
\label{sec:results}

\subsection{Density signatures of the liquid--liquid transition}
\label{sec:density}

\begin{figure}[htbp]
\centering
\includegraphics[trim={0cm 0 0cm 2cm},clip,width=0.62\textwidth]{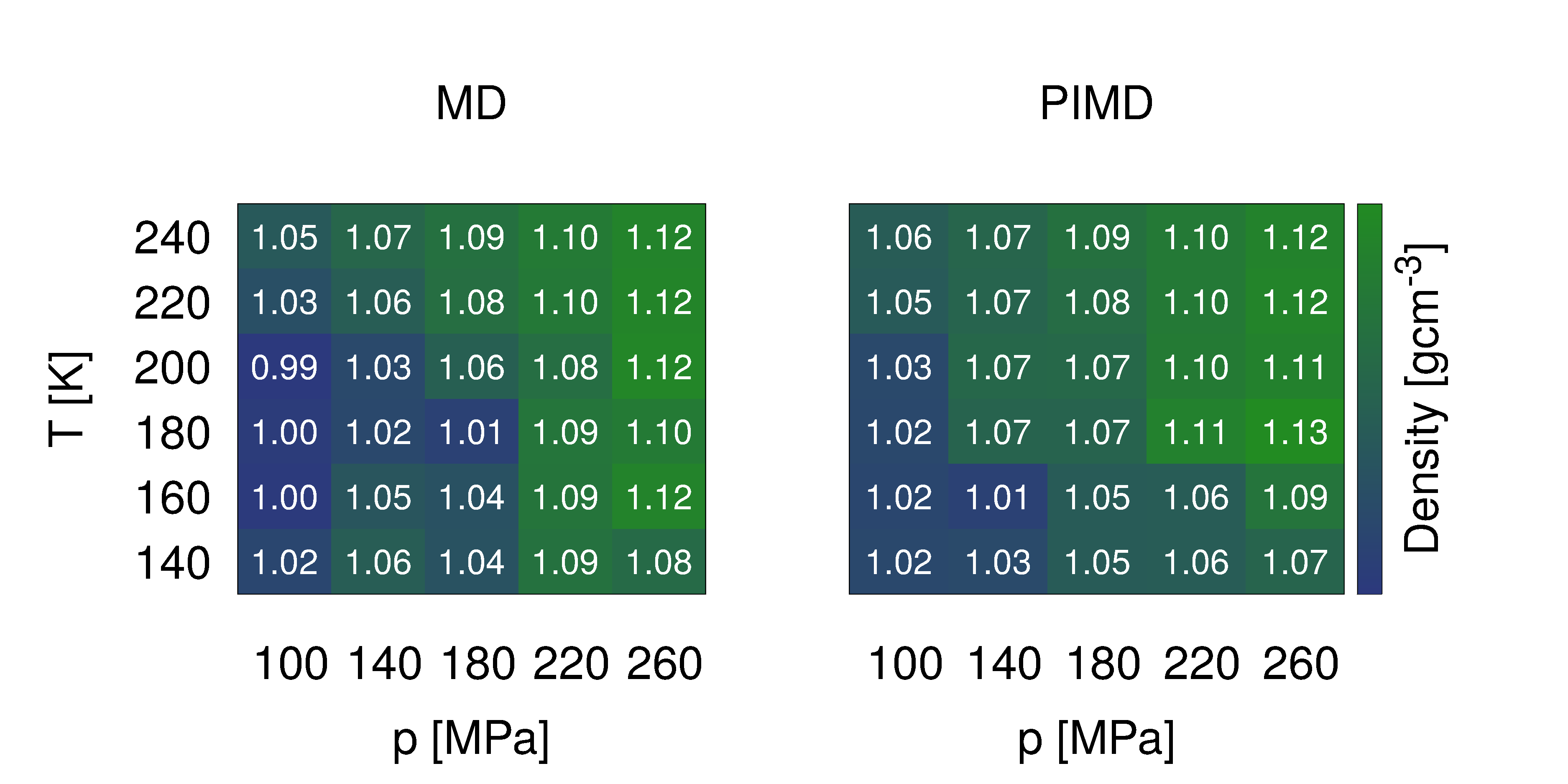}
\includegraphics[trim={0cm 5cm 0cm 0},clip,width=0.31\textwidth]{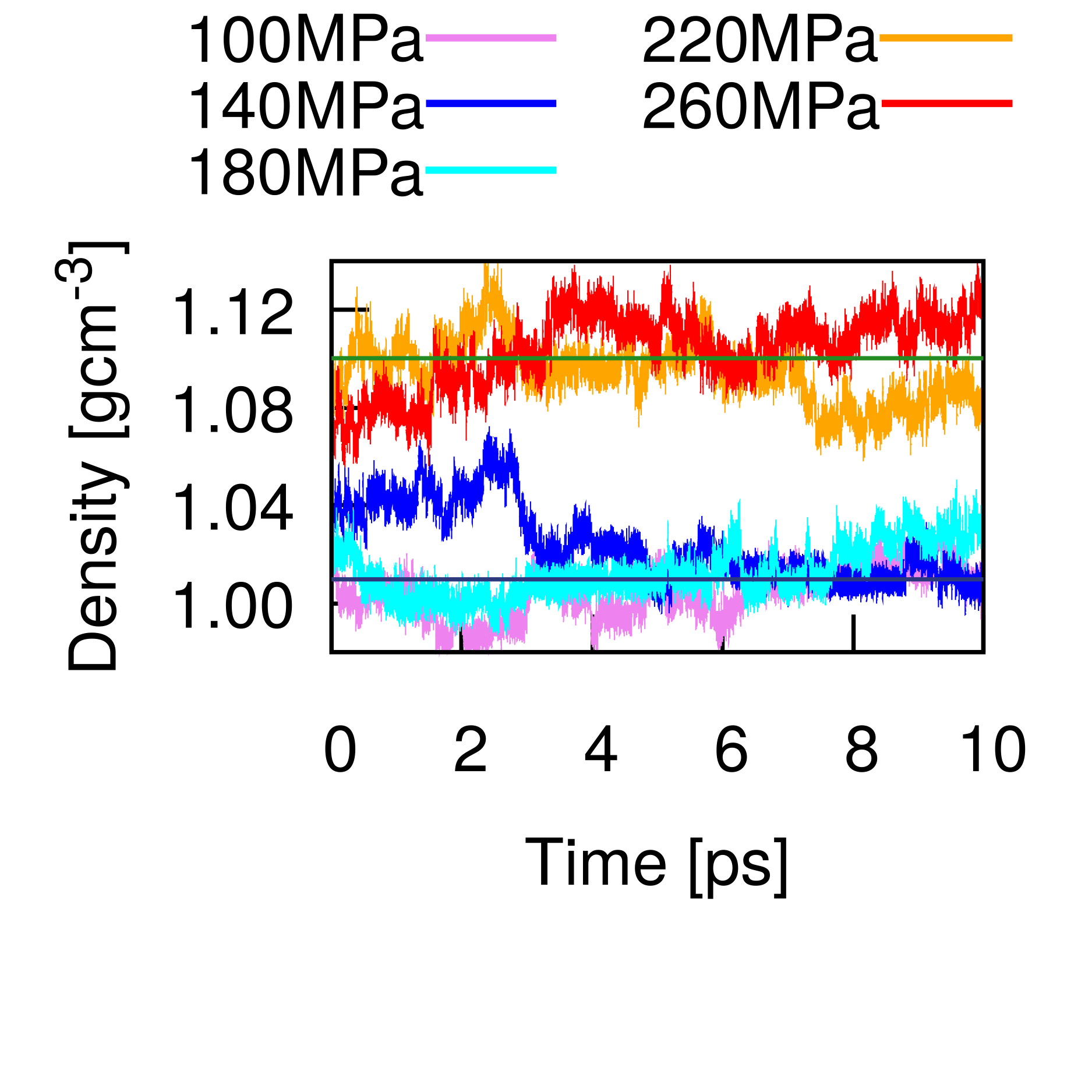}
\caption{Average densities of the MD and PIMD simulations of water around the expected LLCP (left) and density time traces from the classical MD simulations at 180 K for the five pressures considered (right).}
\label{Fig:density}
\end{figure}

Figure~\ref{Fig:density} shows the average densities obtained from classical MD and PIMD as a function of temperature and pressure. The classical simulations display a pronounced density increase at 180 K between 180 and 220 MPa, from approximately 1.0 to 1.1 g cm$^{-3}$. The density time traces at 180 K fluctuate around stable mean values of about 1.01 and 1.10 g cm$^{-3}$ on the low- and high-pressure sides, respectively. This behavior is consistent with a transition between LDL-like and HDL-like liquid states in the classical simulations.

The PIMD simulations do not show an equally sharp density discontinuity over the same grid of thermodynamic states. Instead, the density increases more smoothly with pressure in the low-temperature regime. This difference has two possible interpretations. NQE may shift, weaken, or remove the first-order transition in this model, or the transition may still be present but not resolvable from the present finite set of 10 ns trajectories because density is a slowly converging quantity in deeply supercooled water. The present data therefore support a clear structural influence of NQE, but they are not sufficient to determine a precise LLCP location.

\subsection{Radial distribution functions}
\label{sec:rdf}

\begin{figure}[htbp]
\centering
\includegraphics[trim={1cm 0 1cm 0},clip,width=0.31\textwidth]{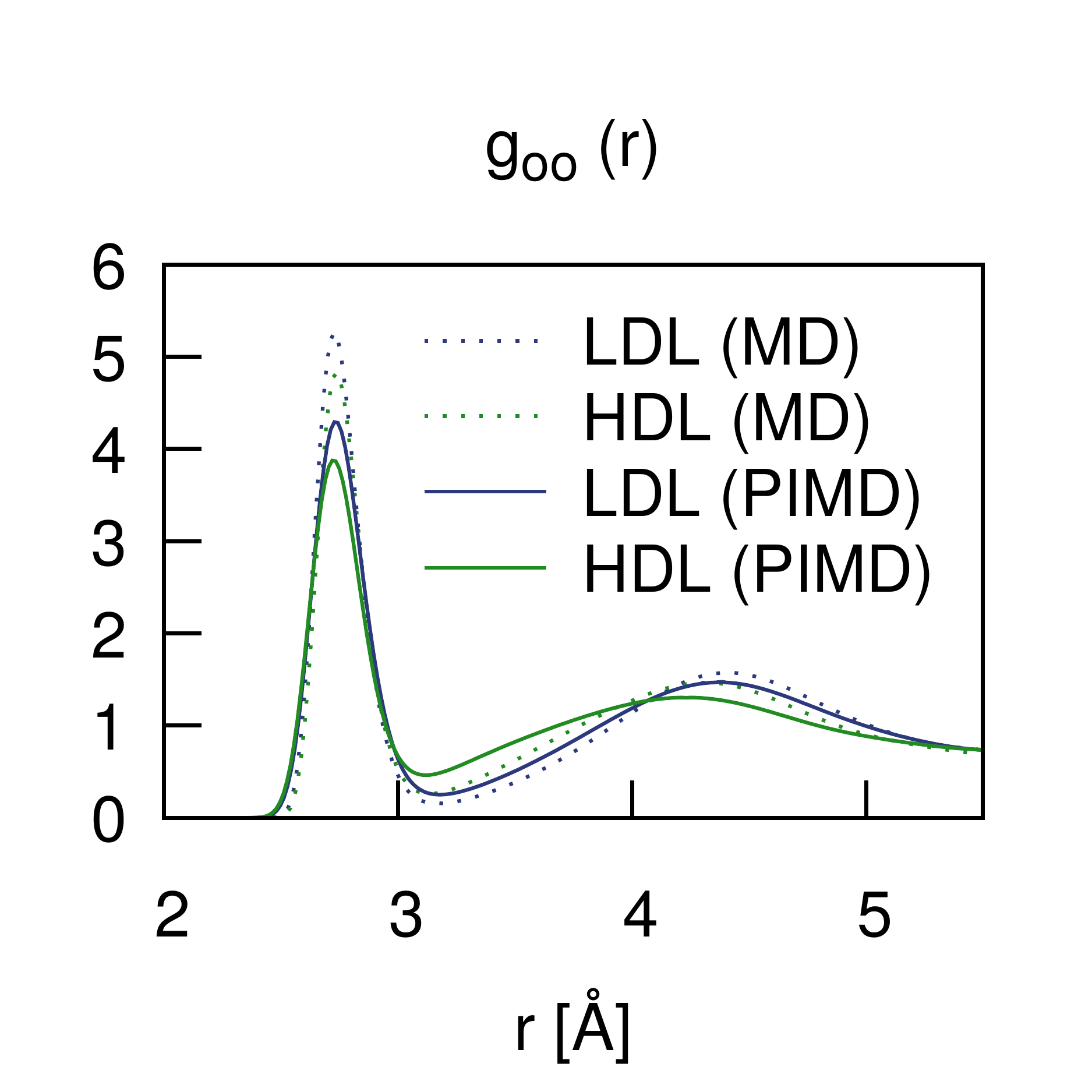}
\includegraphics[trim={1cm 0 1cm 0},clip,width=0.31\textwidth]{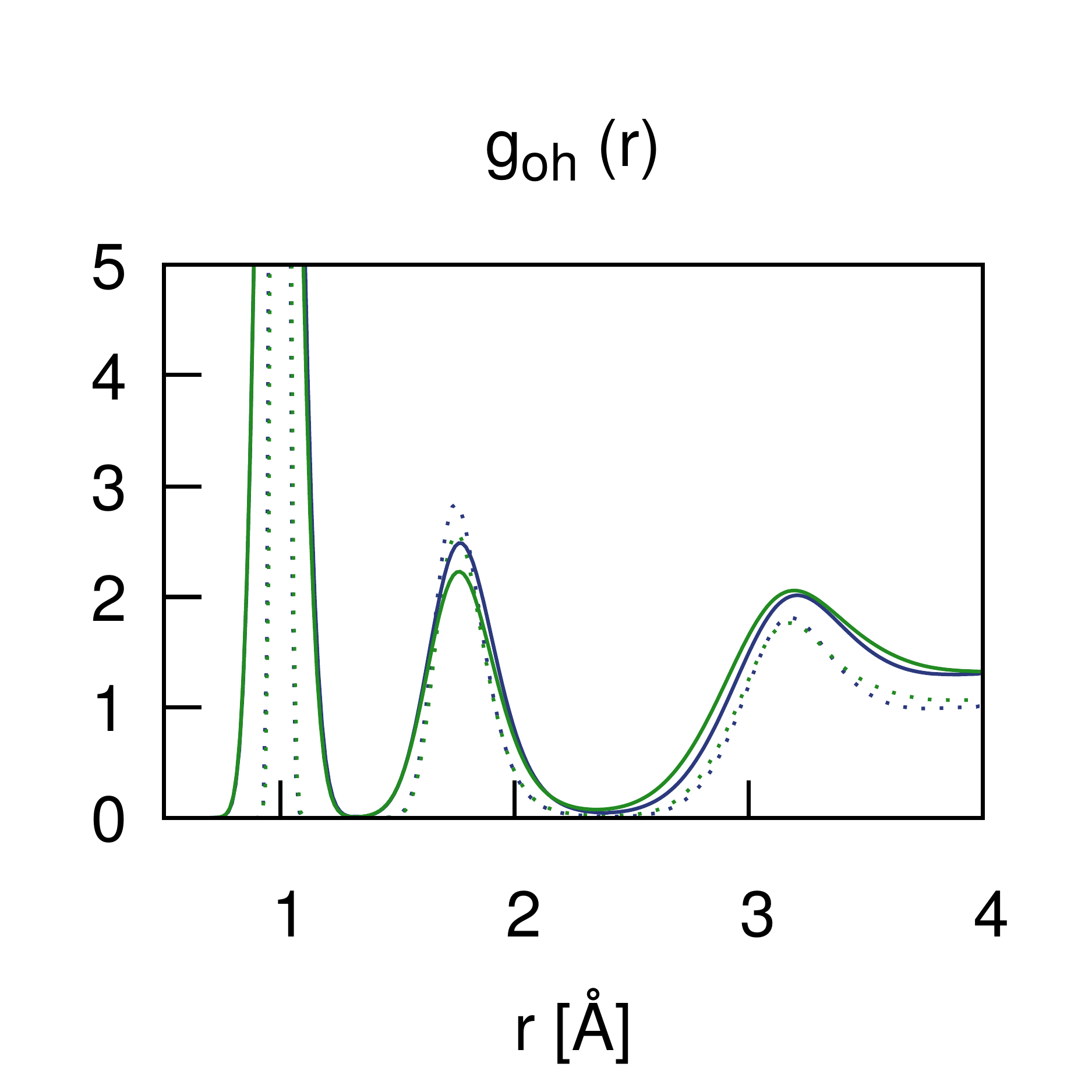}
\includegraphics[trim={1cm 0 1cm 0},clip,width=0.31\textwidth]{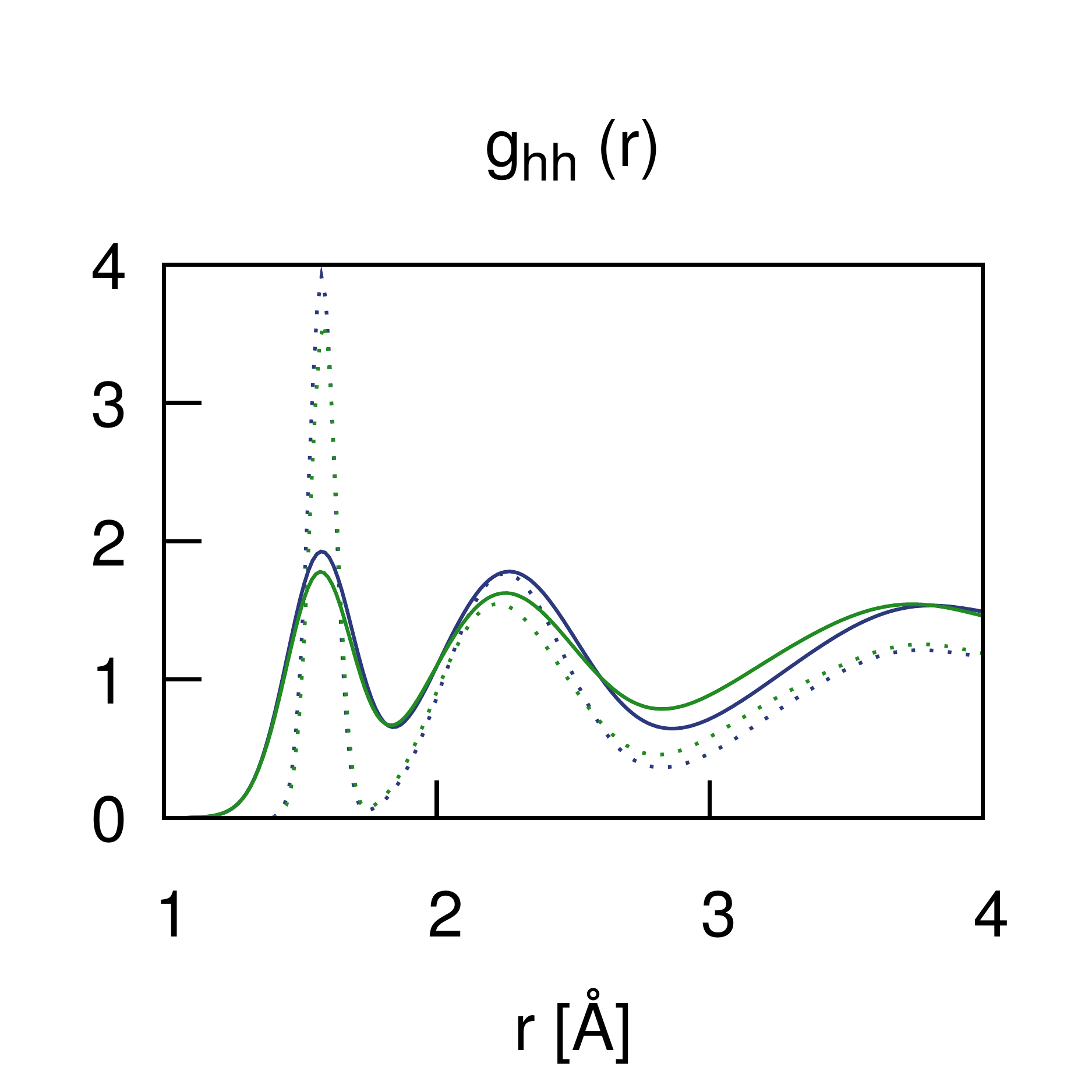}
\caption{Effect of NQE on the radial distribution functions $g_{\mathrm{OO}}(r)$, $g_{\mathrm{OH}}(r)$, and $g_{\mathrm{HH}}(r)$ of LDL-like and HDL-like water at 180 K. The RDFs are compared between PIMD and classical MD.}
\label{Fig:gofr}
\end{figure}

To analyze how NQE change the microscopic structure, we compare RDFs for two representative state points at 180 K. The 100 MPa simulation is used as an LDL-like state and the 220 MPa simulation as an HDL-like state, based on the density jump observed in the classical simulations. Figure~\ref{Fig:gofr} shows the oxygen--oxygen, oxygen--hydrogen, and hydrogen--hydrogen RDFs for classical MD and PIMD.

Including NQE lowers the peak heights, broadens the peaks, and makes the RDFs generally less structured. This broadening is expected from the zero-point motion of the nuclei and is most pronounced in pair correlations involving hydrogen. The oxygen--oxygen RDF is affected more weakly, consistent with the larger mass of oxygen. The HDL-like RDFs are shifted to slightly shorter distances relative to the LDL-like RDFs, reflecting the higher density of the HDL-like state. Overall, the RDFs show that NQE do not merely perturb the hydrogen coordinates locally; they soften the structural order of both LDL-like and HDL-like liquid water in the LLPT region.

\subsection{Bond-order parameters}
\label{sec:bop_results}

\begin{figure}[htbp]
\centering
\includegraphics[trim={1cm 0 3cm 0cm},clip,width=0.62\textwidth]{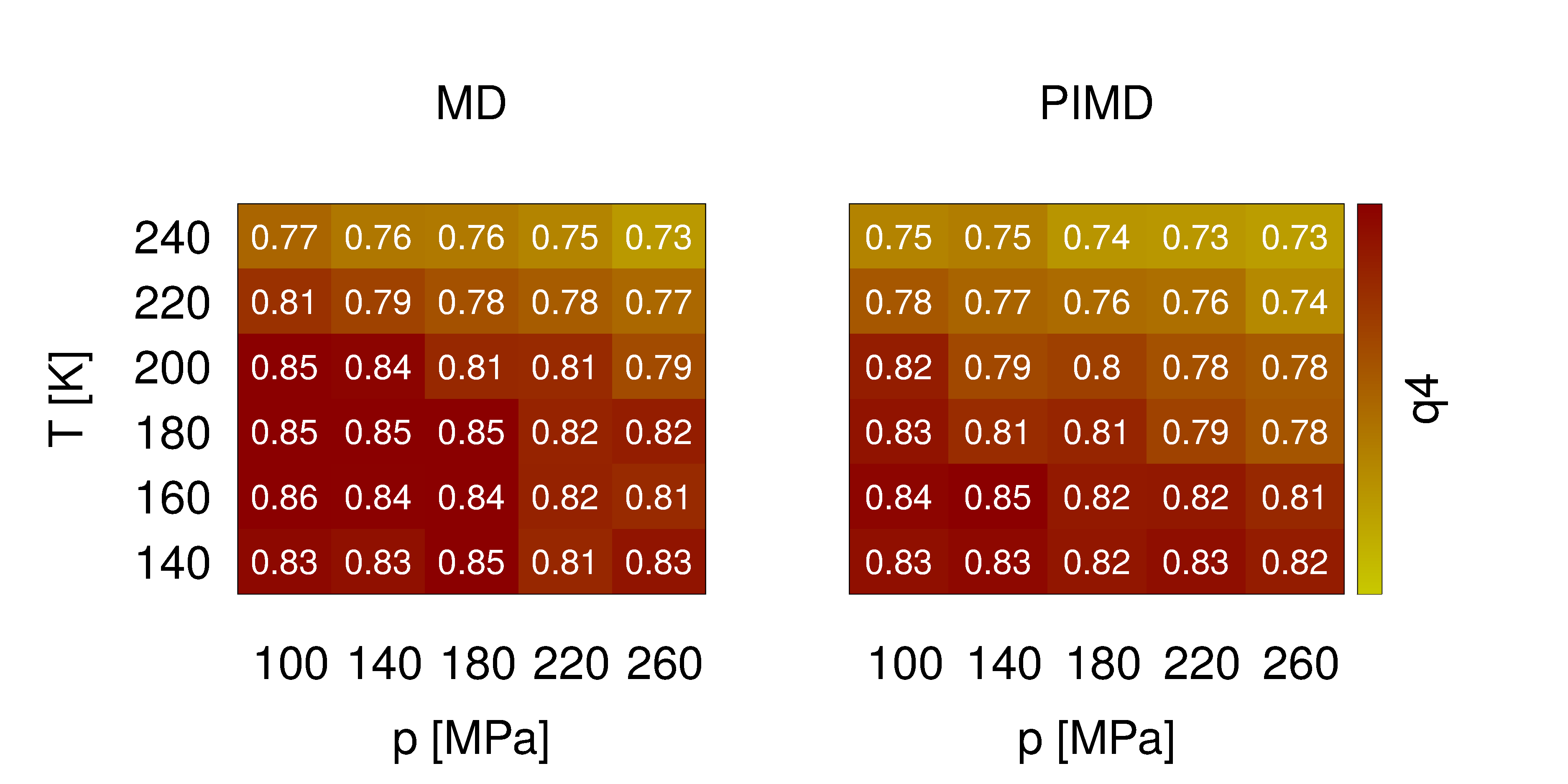}
\includegraphics[trim={1cm 0 0cm 1cm},clip,width=0.31\textwidth]{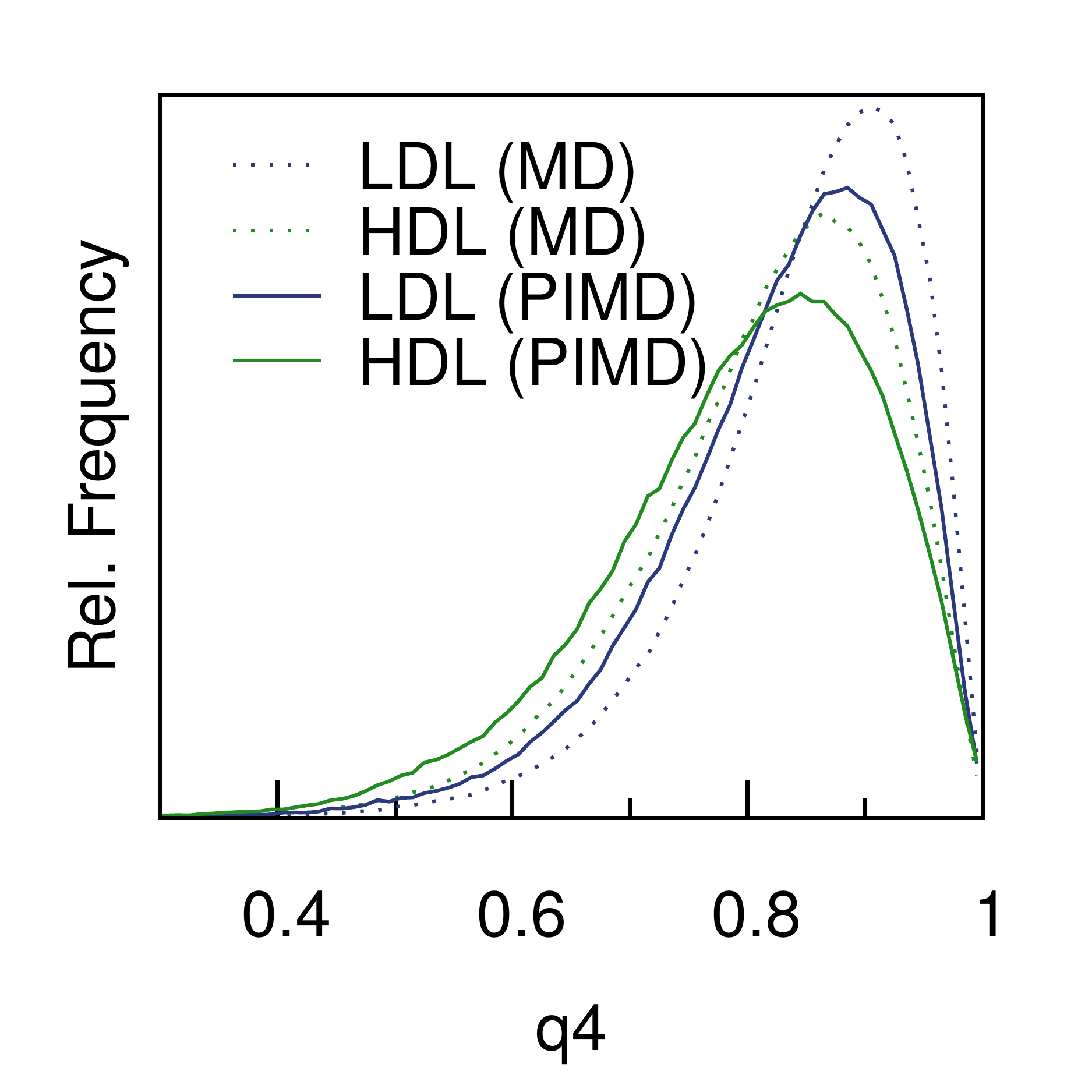} \\
\includegraphics[trim={1cm 0 3cm 0cm},clip,width=0.62\textwidth]{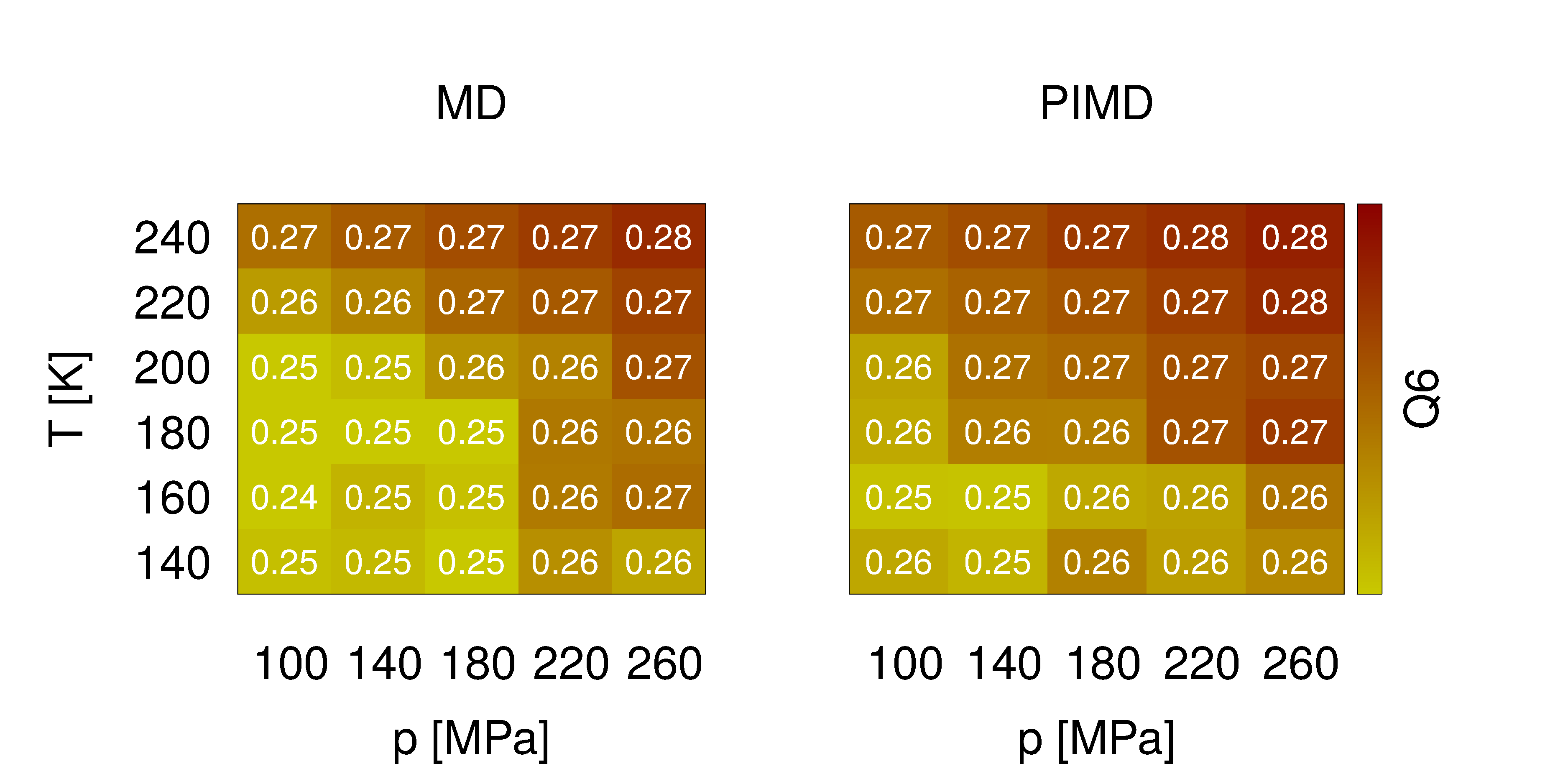}
\includegraphics[trim={1cm 0 0cm 1cm},clip,width=0.31\textwidth]{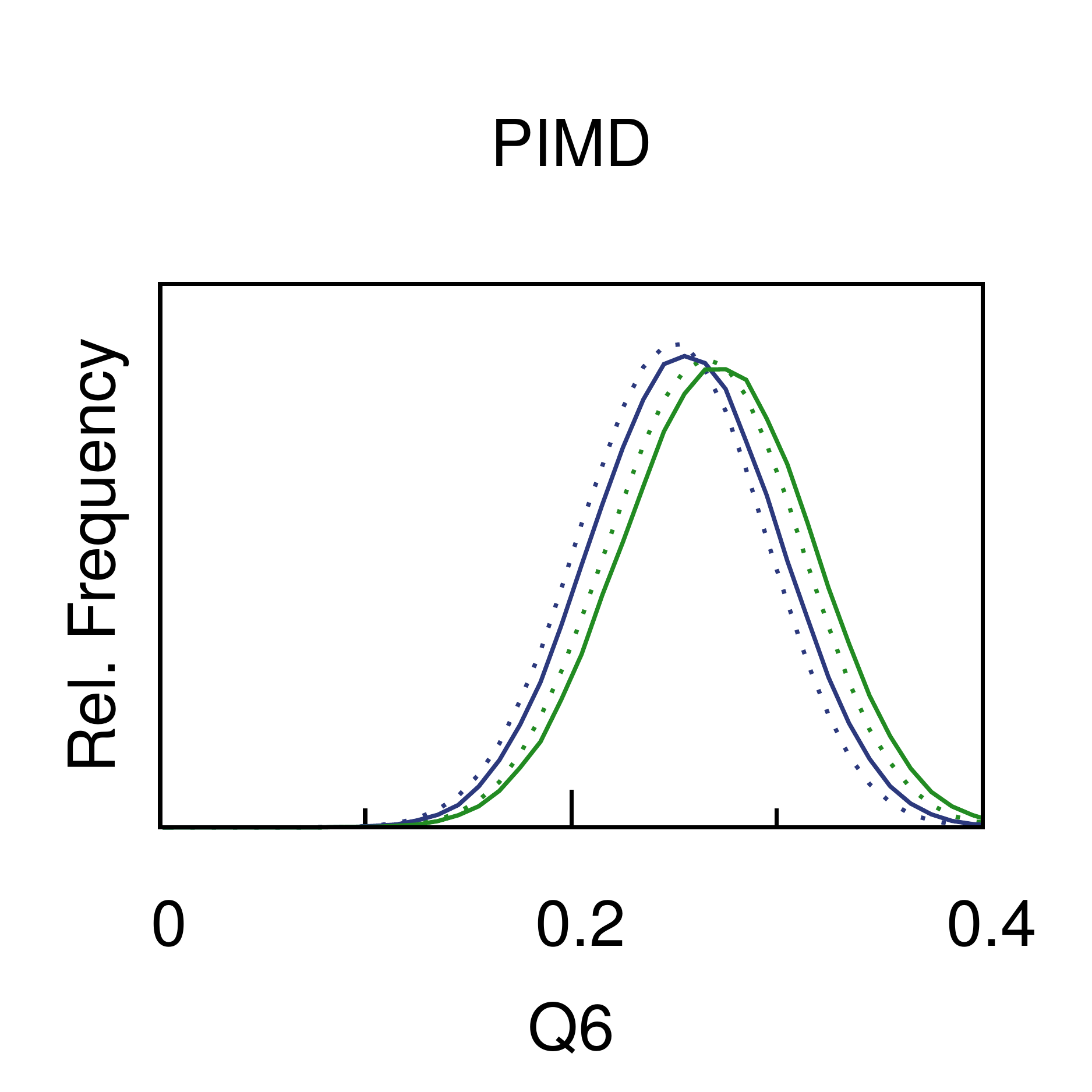}
\caption{Effect of NQE on the tetrahedral order parameter $q_4$ (upper panels) and the Steinhardt parameter $Q_6$ (lower panels). The left panels show averages over the investigated thermodynamic state points, while the right panels show distributions for representative LDL-like and HDL-like states.}
\label{Fig:q4Q6}
\end{figure}

The trends observed in the RDFs are reflected in the bond-order parameters shown in Fig.~\ref{Fig:q4Q6}. The average tetrahedral order parameter $q_4$ lies between 0.75 and 0.83 across the investigated state points. These values exceed reported values of 0.6686 at 298 K and 0.7297 at 260 K for ambient-pressure liquid water \cite{q4_bulk_water}, consistent with the enhanced tetrahedral ordering expected in deeply supercooled water.

Temperature has the strongest influence on $q_4$: lowering the temperature increases tetrahedral order until the trend reaches a maximum or plateau around 180 K. At the representative 180 K state points, HDL-like water is less tetrahedral than LDL-like water. PIMD systematically lowers $q_4$ relative to the corresponding classical simulations, demonstrating that NQE reduce first-shell tetrahedral order. This reduction is also visible in the distributions of local $q_i$ values, which shift toward lower tetrahedrality when quantum nuclei are included.

The Steinhardt $Q_6$ parameter provides complementary information on ordering extending toward the second hydration shell. In contrast to $q_4$, $Q_6$ increases with increasing temperature and pressure in the present simulations. The values remain close to the uncorrelated reference value $Q_6^{\mathrm{ig}}\approx 0.289$ for twelve neighbors, especially at higher temperatures and pressures. Thus, while the first shell remains strongly tetrahedral, the second-shell ordering is far from crystalline. Because $Q_6$ depends on the neighbor definition used in the analysis \cite{Mickel2013}, its absolute value should not be over-interpreted; the robust information is the relative trend across state points and between classical and quantum nuclei.

The $Q_6$ distributions show a small shift to higher values in the PIMD simulations. This NQE-induced shift is weaker than the corresponding decrease in $q_4$, indicating that quantum nuclei most strongly disrupt first-shell tetrahedrality while producing only a modest reorganization of the more extended local environment. Together, the $q_4$ and $Q_6$ results emphasize that NQE affect LDL-like and HDL-like water differently across structural length scales.

These results also bear on structural two-state interpretations of water, in which anomalous thermodynamics are linked to fluctuations between locally HDL-like and LDL-like environments \cite{Nilsson2015}. Our simulations do not rule out such thermodynamic descriptions, but they caution against identifying two sharply separated molecular species from static order parameters alone. The distinction between LDL-like and HDL-like water is sensitive to the structural descriptor: $q_4$ suggests reduced first-shell tetrahedrality upon including NQE, whereas $Q_6$ changes in the opposite direction and remains close to the uncorrelated reference. A related caveat applies to inherent-structure analyses, where energy minimization removes thermal and quantum fluctuations and can sharpen bimodality in local-structure distributions \cite{Wikfeldt2011}. The NQE-induced broadening of pair correlations and reduction of tetrahedral order therefore caution against interpreting static structural order parameters as evidence for sharply separated local LDL- and HDL-like species, particularly when nuclear quantum motion is neglected.

\section{Conclusions}
\label{sec:conclusions}

We have compared classical MD and PIMD simulations of flexible q-TIP4P/F-like water in the deeply supercooled region around the expected LLCP. The classical simulations show a pronounced density change at 180 K between 180 and 220 MPa, consistent with a transition between LDL-like and HDL-like liquid states. In contrast, the PIMD simulations show a smoother pressure dependence over the same state-point grid, indicating that NQE substantially modify the thermodynamic and structural signatures used to identify the LLPT.

The structural analysis shows that NQE broaden the RDFs, reduce first-shell tetrahedrality, and slightly increase the Steinhardt $Q_6$ parameter. These changes occur in both LDL-like and HDL-like states, demonstrating that NQE are not a small correction in the no man's land regime. The present simulations do not determine the LLCP location precisely, but they show that any such determination with water models of this type should include quantum nuclei. Longer trajectories and finite-size analysis will be required to distinguish a shifted LLPT from a genuine smoothing or disappearance of the transition.

\begin{acknowledgments}
The authors have received funding from the European Research Council (ERC) under the European Union's Horizon 2020 research and innovation programme (Grant Agreement No. 716142). The generous allocation of computing time on the FPGA-based supercomputer ``Noctua'' by the Paderborn Center for Parallel Computing (PC2) is gratefully acknowledged.
\end{acknowledgments}

\bibliographystyle{aipnum4-2}
\bibliography{repo}

\end{document}